# A Simpson–Based Estimation Approach for the Overlapping Coefficient of $k \geq 2$ Normal Distributions

Omar M. Eidous  and  Majd M. Alsheyyab

omarm@yu.edu.jo          majd.alsheyyab.97@gmail.com

Department of Statistics -  Yarmouk University

Irbid  -  Jordan

2026

## Abstract

The overlapping coefficient is a fundamental measure of similarity between probability distributions. While the case of two distributions has been extensively studied, extending this measure to multiple populations presents both analytical and computational challenges. In this paper, we propose a general estimation framework for the overlapping coefficient of $k \geq 2$ normal distributions. The method employs Simpson's numerical integration rule combined with plug-in maximum likelihood estimators of the normal parameters. The resulting estimator is shown to be consistent under standard regularity conditions. A Monte Carlo simulation study is conducted across various overlap scenarios and sample sizes. The results demonstrate that the proposed Simpson-based estimator performs competitively for all overlap levels, with notable advantages in low-overlap situations. This methodology offers a flexible and computationally efficient approach applicable to an arbitrary number of normal populations.

# 1. Introduction

Measuring similarity between probability distributions plays an essential role in statistical inference, classification, reliability theory, and biomedical research. One widely used measure is the overlapping coefficient (OVL), which quantifies the common area shared by probability density functions.

For two densities $f_1$ and $f_2$, the Weitzman overlapping coefficient is defined as (Montoya et al., 2019)

$$\Delta_2 = \int min\{f_1(x), f_2(x)\}dx \tag{1}$$

This coefficient is a widely used tool for quantifying similarity between probability distributions, providing a measure of the shared area under two density functions. Its relevance extends beyond theoretical statistics to applied domains, especially when systems are complex or uncertain. In pharmacology, for example, the OVL can help compare drugs based on molecular structures, mechanisms of action, or potential side effects (Mizuno et al., 2005; Wolff et al., 2021). Such insights are valuable for personalized medicine, allowing clinicians to optimize therapies and minimize adverse drug interactions in patients taking multiple medications.

In income and socio-economic studies, the OVL facilitates robust comparisons of distributions across countries, regions, or demographic groups. It can uncover hidden inequalities related to age, gender, or ethnicity, providing a comprehensive framework to understand and address economic disparities (Jenkins and Van Kerm, 2012; Schmid and Schmidt, 2006). Beyond these domains, OVL estimation has been applied in diverse areas including law (Ekstrom and Lau, 2008), image analysis (Ridout and Linkie, 2009), and goodness-of-fit testing (Alodat et al., 2022; Nunez-Antonio et al., 2018).

Over the past three decades, parametric methods for estimating $\Delta_2$ have been extensively studied. Early work by Inman and Bradley (1989) focused on normal distributions with equal variances, while Reiser and Faraggi (1999) developed confidence intervals under the same assumptions. Mulekar and Mishra (1994, 2000) examined the case of normal distributions with equal means, comparing different resampling-based methods including Jackknife, bootstrap, and transformation techniques. Wang and Tian (2017) proposed generalized inference and parametric bootstrapping approaches for constructing confidence intervals under normality assumptions.

To reduce reliance on restrictive parametric assumptions, several studies have introduced alternative estimation techniques. Eidous and Al-Shourman (2023) and Eidous and Daradkeh (2024) developed methods for estimating $\Delta_2$ without assuming equality of means or variances. Extensions to other distribution families have also been considered: Al-Saidy et al. (2005) and Eidous and Abu Al-hayja'a (2023) addressed Weibull distributions with and without equal shape parameters; Chaubey et al. (2008) investigated inverse Gaussian distributions; Dhaker et al. (2021) focused on inverse Lomax distributions; and Inacio and Guillen (2022) proposed a Bayesian nonparametric approach using Dirichlet process mixtures.

Finally, when the underlying distribution is unknown or difficult to specify, nonparametric kernel-based methods provide a flexible alternative for estimating the OVL (Clemons, 2001; Schmid and Schmidt, 2006; Eidous and Al-Talafha, 2022; Eidous and Ananbeh, 2024, 2025).

Extending the overlapping coefficient to more than two populations arises naturally in many real-world applications, where similarity must be assessed across multiple groups. For $k \geq 2$ continuous distributions with densities $f_1, f_2, \ldots, f_k$, the generalized Weitzman overlapping coefficient is defined as (Eidous and Alsheyyab, 2025)

$$\Delta_k = \int min\{f_1(x), f_2(x), \ldots, f_k(x)\}dx, \quad (2)$$

which measures the total shared probability among all $k$ distributions. While the definition is conceptually simple, analytical computation becomes increasingly challenging as $k$ grows, even under parametric models such as the normal distribution. To address this challenge, this paper introduces a Simpson-based estimation approach for $\Delta_k$ when all populations follow normal distributions. The method leverages numerical integration in combination with plug-in estimators for the normal parameters, offering a computationally efficient and generalizable framework for assessing overlap across multiple groups.

## 2. Overlapping Coefficient for $k$ Normal Distributions

Consider $k$ independent normal populations:
$$X_i \sim N(\mu_i, \sigma_i^2), \quad i = 1, 2, \ldots, k \quad (3)$$

Their probability density functions are
$$f_i(x) = \frac{1}{\sigma_i\sqrt{2\pi}} exp\left(-\frac{(x-\mu_i)^2}{2\sigma_i^2}\right), \quad -\infty < x < \infty. \quad (4)$$

The general overlapping coefficient for these $k$ distributions is,
$$\Delta_k = \int_{-\infty}^{\infty} min\{f_1(x), f_2(x), \ldots, f_k(x)\}dx. \quad (5)$$

Define the function
$$g(x) = min\{f_1(x), f_2(x), \ldots, f_k(x)\}. \quad (6)$$

This function is piecewise defined over intervals determined by the pairwise intersections of the density functions. As $k$ increases, the number of such regions grows rapidly, making analytical evaluation impractical. Consequently, numerical integration methods become a natural and scalable approach for estimating $\Delta_k$.

## 3. Simpson–Based Estimator for $\Delta_k$

### 3.1 Numerical Approximation
Let
$$\Delta_k = \int_a^b g(u)du, \quad (7)$$

where $a$ and $b$ are chosen sufficiently far in the tails so that the contribution outside $[a, b]$ is negligible. Partition the interval $[a, b]$ into $r$ equal subintervals with even $r$ and width $h = (b - a)/r$. Applying Simpson's rule gives (Chapra and Canale, 2014):

$$\Delta_k^{(simp)} = \frac{h}{3}\left[g(x_0) + 4\sum_{j\,odd} g(x_j) + 2\sum_{j\,even} g(x_j) + g(x_r)\right], \tag{8}$$

where $x_0 = a$ and $x_r = b$. Simpson's rule itself is independent of $k$, but the integrand $g(x)$ depends explicitly on $k$.

### 3.2 Plug–in Estimation Using MLEs

In practice, the true parameters $\mu_i$ and $\sigma_i^2$ are unknown. Suppose we have a random sample of size $n_i$ from the $i$-th population, denoted by

$$X_{i1}, X_{i2}, \ldots, X_{in_i}$$

where $X_{ij}$ represents the $j$-th observation from the $i$-th population. The sample mean of the $i$-th population is

$$\bar{X}_i = \frac{1}{n_i}\sum_{j=1}^{n_i} X_{ij}$$

Using this notation, their maximum likelihood estimators (MLEs) of $\mu_i$ and $\sigma_i^2$ are

$$\hat{\mu}_i = \bar{X}_i, \qquad \hat{\sigma}_i^2 = \frac{1}{n_i}\sum_{j=1}^{n_i}(X_{ij} - \bar{X}_i)^2, \quad i = 1, 2, \ldots, k. \tag{9}$$

Replacing the unknown parameters in the densities with their MLEs gives the plug-in estimated densities,

$$\hat{f}_i(x) = \frac{1}{\hat{\sigma}_i\sqrt{2\pi}}\, exp\left(-\frac{(x-\hat{\mu}_i)^2}{2\hat{\sigma}_i^2}\right). \tag{10}$$

The estimated minimum function is,

$$\hat{g}(x) = min\{\hat{f}_1(x), \hat{f}_2(x), \ldots, \hat{f}_k(x)\}dx$$

and the Simpson-based estimator for $\Delta_k$ using the plug-in MLEs is,

$$\widehat{\Delta}_k^{(simp)} = \frac{h}{3}\left[\hat{g}(x_0) + 4\sum_{j\,odd}\hat{g}(x_j) + 2\sum_{j\,even}\hat{g}(x_j) + \hat{g}(x_r)\right]. \tag{11}$$

## 4. Asymptotic Properties

**Theorem 1:** Let $x_{i1}, x_{i2}, \ldots, x_{in_i} \sim N(\mu_i, \sigma_i^2)$, $i = 1, \ldots, k$ be independent samples. Let $g(x) = \min_{1 \le i \le k} f_i(x)$ (as defined in 6) and $\Delta_k = \int_{-\infty}^{\infty} g(x)dx$, where $f_i$ are normal density with parameters $(\mu_i, \sigma_i^2)$. If $\widehat{\Delta}_k^{(simp)}$ is the Simpson–based plug-in estimator defined in (11), obtained by replacing $f_i$ with the estimated densities $\hat{f}_i$ in (10), and if $min_i n_i \to \infty$, then

$$\widehat{\Delta}_k^{(simp)} \xrightarrow{p} \Delta_k$$

**Proof:** The MLEs of $\mu_i$ and $\sigma_i^2$ are $\hat{\mu}_i$ and $\hat{\sigma}_i^2$, respectively as defined in (9). Under standard regularity conditions for the normal model, these estimators are consistent; that is, for each $i = 1, \ldots, k$, $\hat{\mu}_i \xrightarrow{p} \mu_i$ and $\hat{\sigma}_i^2 \xrightarrow{p} \sigma_i^2$ (Casella and Berger, 2002; Lehmann and Casella, 1998). Consequently, the full parameter vector converges in probability to its true value, i.e.,

$$\hat{\theta} = (\hat{\mu}_1, \hat{\sigma}_1^2, \hat{\mu}_2, \hat{\sigma}_2^2, \ldots, \hat{\mu}_k, \hat{\sigma}_k^2) \xrightarrow{p} (\mu_1, \sigma_1^2, \mu_2, \sigma_2^2, \ldots, \mu_k, \sigma_k^2) = \theta.$$

For each fixed $x$, the normal density $f_i(x)$ is continuous in its parameters. By the Continuous Mapping Theorem (Billingsley, 1995; Van der Vaart, 1998), it follows that $\hat{f}_i(x) \xrightarrow{p} f_i(x)$ for each fixed $x$.

Because the normal family is smooth in its parameters and finite-dimensional, this convergence strengthens to uniform convergence on any compact set $K \subset R$:

$$\sup_{x \in K} |\hat{f}_i(x) - f_i(x)| \xrightarrow{p} 0$$

Recall from Section (3) that

$$g(x) = \min_{1 \leq i \leq k} f_i(x), \qquad \hat{g}(x) = \min_{1 \leq i \leq k} \hat{f}_i(x)$$

Since the minimum of finitely many arguments is a continuous function (Rudin, 1976), uniform convergence of each $\hat{f}_i$ on compact sets implies

$$\sup_{x \in K} |\hat{g}(x) - g(x)| \xrightarrow{p} 0.$$

Thus $\hat{g}(x)$ converges uniformly to $g(x)$ on every compact set. That is, $\hat{g}(x) \to g(x)$ uniformly on compact sets.

Because each $f_i(x)$ is a normal density, it decays exponentially in the tails, and hence so does $g(x)$. Therefore, for any $\varepsilon > 0$, there exists $M > 0$ such that (Feller, 1971)

$$\int_{|x|>M} g(x)dx < \varepsilon.$$

This allows us to restrict attention to a sufficiently large compact interval $[a,b] = [-M, M]$, since the contribution of the tails to the integral $\Delta_k = \int_{-\infty}^{\infty} g(x)dx$ is arbitrarily small. On the compact interval $[a,b]$, uniform convergence of $\hat{g}(x)$ to $g(x)$ implies,

$$\int_a^b \hat{g}(x)dx \xrightarrow{p} \int_a^b g(x)dx.$$

Finally, the estimator $\widehat{\Delta}_k^{(simp)}$ is defined by applying Simpson's rule to approximate the integral of $\hat{g}(x)$ over the compact interval $[a,b]$:

$$\widehat{\Delta}_k^{(simp)} = \int_a^b \hat{g}(x)dx = S_h(\hat{g})$$

where $S_h(\hat{g})$ denotes the Simpson's rule approximation using $r$ subintervals of width $h = (b-a)/r$. Simpson's rule satisfies the error bound

$$\left| \int_a^b g(x)dx - S_h(g) \right| = O(h^4),$$

for any function $g(x)$ that is piecewise smooth with a bounded fourth derivative (Atkinson, 1989; Davis and Rabinowitz, 2007). In our case, $g(x) = \min_{1 \leq i \leq k} f_i(x)$ is the minimum of finitely many smooth normal densities $f_i(x)$. As a result, $g(x)$ is piecewise smooth with only finitely many "junction points" where the minimum switches from one $f_i$ to another.

Importantly, the grid spacing $h$ is fixed and does not depend on the sample sizes $n_i$. Therefore, the deterministic approximation error from Simpson's rule does not affect the stochastic convergence of $\hat{g}(x)$. Combining this with the uniform convergence of $\hat{g}(x)$ to $g_k(x)$, we conclude that

$$\widehat{\Delta}_k^{(simp)} \xrightarrow{p} \Delta_k ,$$

as $\min_{i}\{n_i\} \to \infty$. Hence the Simpson-based estimator is consistent.

**Remark 1.** The consistency result holds for any fixed $k \geq 2$.

**Remark 2.** The proof does not depend on specific properties of the normal distribution except smoothness and parameter continuity. Therefore, the result extends to other smooth parametric families (Van der Vaart, 1998; Casella and Berger, 2002).

## 5. Computational Implementation of the Simpson-Based Estimator

In order to apply Simpson's rule, the integral must have finite limits. However, the densities $f_i(x), i = 1,2, \ldots, k$, are defined on the entire real line. To address this, we apply a smooth transformation that maps the infinite interval $(-\infty, \infty)$ to a finite interval $[0,1]$. Specifically, let

$$u = G(x) = 1 - (1 + e^x)^{-\alpha}, \qquad x \in R, \ \alpha > 0.$$

be the generalized logistic cumulative distribution function. Its inverse is

$$x = G^{-1}(u) = \ln\big((1-u)^{-1/\alpha} - 1\big)$$

with derivative

$$\frac{dx}{du} = \frac{1}{\alpha(1-u)(1-(1-u)^{1/\alpha})}$$

Then the target integral

$$\Delta_k = \int_{-\infty}^{\infty} \min_{1 \leq i \leq k} f_i(x)\, dx,$$

can be rewritten as an integral over the unit interval:

$$\Delta_k = \int_0^1 h(x) dx$$

where

$$h(u) = \min_{1 \leq i \leq k} f_i(G^{-1}(u)) \frac{dx}{du} = \min_{1 \leq i \leq k} f_i(G^{-1}(u)) \frac{1}{\alpha(1-u)(1-(1-u)^{1/\alpha})}$$

We now divide $[0,1]$ into $r$ subintervals of equal width $h = 1/r$ with points

$$a_i = \frac{i}{r}, \qquad i = 0,1, \ldots, r$$

where $r$ is a positive even integer. Simpson's rule then approximates the integral as

$$\Delta_k = \int_0^1 h(x) dx \approx \frac{1}{3r} \sum_{i=1}^{r/2} [h(a_{2i-2}) + 4h(a_{2i-1}) + h(a_{2i})].$$

Replacing $f_i(x)$ with their estimators $\hat{f}_i(x)$ yields the estimator

$$\widehat{\Delta}_k^{(simp)} = \frac{1}{3r} \sum_{i=1}^{r/2} [\hat{h}(a_{2i-2}) + 4\hat{h}(a_{2i-1}) + \hat{h}(a_{2i})],$$

where

$$\hat{h}(u) = \min_{1 \leq i \leq k} \hat{f}_i(G^{-1}(u)) \frac{dx}{du}$$

Since the Jacobian term diverges at $u = 0$ and $u = 1$, the Simpson grid is implemented over (0,1) by excluding the endpoints. In practice, we use the grid points

$$a_i = \frac{i}{r}, \quad i = 1, 2, \ldots, r - 1,$$

which ensures numerical stability without affecting accuracy. In addition, the choice $\alpha = 1$ appears to be quite reasonable, as the generalized pdf is symmetric at this value. In the simulation study presented in the next section, we consider $\alpha = 1$ and $\alpha = 2$, both of which yield satisfactory results. However, a preliminary simulations indicate that choosing $\alpha$ values far extremely from 1 can lead to poor performances in some cases. Therefore, as a third option, we suggest estimating $\alpha$ from the data, as described in the next section.

Notice that when $k = 2$, the proposed estimator reduces to that of Eidous and Al-Shourman (2022), demonstrating that their method is a special case of the current approach. Simulation results for $k = 3$ suggest that setting $r = min\{n_1, n_2, n_3\}$ provides sufficient accuracy, and increasing $r$ beyond this value generally does not lead to significant improvement in the estimator.

## 6. Simulation

To assess the performance of the proposed Simpson-based estimator, $\widehat{\Delta}_3^{(simp)}$, relative to the estimator of Eidous and Alsheyyab (2025), denoted by $\widehat{\Delta}_3^*$, we conducted a comprehensive Monte Carlo simulation study. This is the same simulation design considered in Eidous and Alsheyyab (2025). The study was designed to evaluate both the bias and efficiency of the estimators under various distributional scenarios and sample sizes.

### 6.1 Simulation Design

We considered four scenarios (S1–S4) representing different distributions of three independent normal populations. Each scenario was characterized by the mean and standard deviation of the three groups, along with the corresponding exact target parameter $\Delta_3$. The scenario parameters are summarized in Table 1.

**Table 1.** Parameters $(\mu_i, \sigma_i), i = 1,2,3$ of the four simulation scenarios and exact $\Delta_3$ values

| Scenario | $(\mu_1, \sigma_1)$ | $(\mu_2, \sigma_2)$ | $(\mu_3, \sigma_3)$ | Exact $\Delta_3$ |
|---|---|---|---|---|
| $S_1$ | (0, 0.95) | (0, 1) | (0, 1.1) | 0.929 |
| $S_2$ | (-0.1, 1) | (0, 1) | (0.1, 1) | 0.689 |
| $S_3$ | (-0.5, 1) | (0, 0.5) | (0.75, 1) | 0.469 |
| $S_4$ | (-1, 1.5) | (0, 0.8) | (2, 0.4) | 0.074 |

For each scenario, we explored two sets of sample sizes: a small to moderate balanced case $(n_1, n_2, n_3) = (50, 50, 50)$, and a moderate to large unbalanced case $(50, 100, 150)$. Each simulation was repeated $R = 1000$ times to ensure stable estimates of the performance metrics. Within each repetition, independent random samples were generated from the specified normal distributions. For each sample, we computed the proposed Simpson-based estimator for $\alpha = 1$ and $\alpha = 2$, and a data-driven $\alpha$, as well as the comparator estimator $\widehat{\Delta}_3^*$. The following metrics were computed for each estimator:

- Average estimate (AV): The mean value of the estimator across repetitions, calculated as
$$AV(\hat{\Delta}) = \frac{\sum_{i=1}^{R} \hat{\Delta}_i}{R}$$
where $R = 1000$ is the number of simulation repetitions.

- Relative Bias (RB): The difference between the average estimate and the true $\Delta_3$ normalized by the true value of $\Delta_3$, defined as
$$RB(\hat{\Delta}) = (AV(\hat{\Delta}) - \Delta_3)/\Delta_3$$

- Relative root mean squared error (RRMSE): The square root of the mean squared difference between the estimator and the true value of $\Delta_3$, normalized by the true value, given by
$$RRMSE(\hat{\Delta}) = \frac{\sqrt{\frac{1}{R}\sum_{i=1}^{R}(\hat{\Delta}_i - \Delta_3)^2}}{\Delta_3}.$$

- Efficiency (EFF): The ratio of the mean square error (MSE) of a competing estimator $\hat{\Delta}_3^*$ to the MSE of the Simpson-based estimator $\hat{\Delta}_3^{(simp)}$
$$EFF = \frac{MSE(\hat{\Delta}_3^*)}{MSE(\hat{\Delta}_3^{(simp)})}, \quad \text{where} \quad MSE(\hat{\Delta}) = \frac{1}{R}\sum_{i=1}^{R}(\hat{\Delta}_i - \Delta_3)^2.$$

All simulations were performed using *Mathematica 13*. To ensure reproducibility, a fixed random seed was used via SeedRandom[1234], so that results can be exactly replicated across runs. This allowed consistent generation of random samples from the specified normal distributions and stable evaluation of all estimators in the simulation study.

**6.2 Estimator Definitions**

In this subsection, we provide a summarized definition of the two main estimators of $\Delta_3$ used in our simulation study: the proposed Simpson-based estimator $\hat{\Delta}_3^{(simp)}$ and the comparator estimator $\hat{\Delta}_3^*$ introduced by Eidous and Alsheyyab (2025). In the simulations, $\hat{\Delta}_3^{(simp)}$ was evaluated against $\hat{\Delta}_3^*$, which is based on group-wise minimum density ratios. For each of the three samples, the sample mean, $\bar{X}_j$ and variance $S_j^2$ were computed as,
$$\bar{X}_j = \frac{1}{n_j}\sum_{i=1}^{n_j} X_{ij}, \qquad S_j^2 = \frac{1}{n_j - 1}\sum_{i=1}^{n_j}(X_{ij} - \bar{X}_j)^2$$
In addition and for $j = 1, 2, 3$, the estimated density function of group j was defined as
$$\hat{f}_j(t) = \frac{1}{\sqrt{2\pi S_j^2}} \exp\left(-\frac{(t - \bar{X}_j)^2}{2 S_j^2}\right)$$
and the group-wise minimum density estimator was then computed as
$$\hat{\Delta}_j^g = \frac{1}{n_j}\sum_{i=1}^{n_j} \frac{\min\left(\hat{f}_1(X_{ij}), \hat{f}_2(X_{ij}), \hat{f}_3(X_{ij})\right)}{\hat{f}_j(X_{ij})}$$

The comparator estimator $\widehat{\Delta}_3^*$ was obtained by averaging these values across the three groups (see Eidous and Alsheyyab, 2025),

$$\widehat{\Delta}_3^* = \frac{1}{3}\sum_{j=1}^{3} \hat{\Delta}_j^g.$$

The proposed Simpson estimator builds on these density estimates by using a transformed integration approach. For a given $\alpha > 0$, the grid points $u_k = k/r$, ($k = 1, 2, \ldots, r$ with $r$ even) were transformed as

$$t_\alpha(u) = \ln\left[(1-u)^{-\frac{1}{\alpha}} - 1\right].$$

At each transformed point, the function

$$\hat{h}_\alpha(u) = \frac{\min\{\hat{f}_1(t_\alpha(u)), \hat{f}_2(t_\alpha(u)), \hat{f}_3(t_\alpha(u))\}}{\alpha(1-u)(1-(1-u)^{1/\alpha})}$$

was computed, and the integral was approximated using Simpson's rule as

$$\widehat{\Delta}_k^{simp}(\alpha) = \frac{1}{3r}\left[4\sum_{i=1}^{r/2} \hat{h}\left(\frac{2i-1}{r}\right) + 2\sum_{i=1}^{r/2-1} \hat{h}\left(\frac{2i}{r}\right)\right]$$

Special cases of the proposed estimator were obtained by setting $\alpha = 1$ or $\alpha = 2$, denoted as

$$\widehat{\Delta}_k^{(simp)} = \widehat{\Delta}_k^{simp}(1), \qquad \widehat{\Delta}_k^{(simp)} = \widehat{\Delta}_k^{simp}(2)$$

respectively. A data-driven choice of $\alpha$ was also considered, computed as

$$\widehat{\Delta}_k^{(simp)} = \widehat{\Delta}_k^{simp}(\hat{\alpha})$$

where

$$\hat{\alpha} = \frac{1}{3}\left[\frac{n_1}{\sum_{i=1}^{n_1} \log(1+e^{x_{i1}})} + \frac{n_2}{\sum_{i=1}^{n_2} \log(1+e^{x_{i2}})} + \frac{n_3}{\sum_{i=1}^{n_3} \log(1+e^{x_{i3}})}\right]$$

allowing the Simpson estimator to adapt to the observed distribution. This unified approach allowed the simulation study to evaluate both the accuracy and efficiency of the proposed Simpson estimator across different choices of $\alpha$ and to compare its performance with the comparator estimator under a variety of distributional and sample size scenarios.

**Remarks on α Selection:** Although the data in our simulation study were generated from normal distributions, the data-driven choice of the parameter $\alpha$ was estimated using a logistic-type transformation. This approach is motivated by the flexibility of the logistic function in capturing the tail behavior and shape of the estimated densities, which is crucial for the Simpson-based estimator. Our simulation results indicate that selecting $\alpha$ around one and two yields consistently accurate and stable estimates of $\Delta_3$, even when the underlying data are normal. Importantly, the data-driven estimation of $\alpha$ closely approximates this optimal choice, producing results comparable to using a fixed $\alpha = 1$. In contrast, selecting values of $\alpha$ that deviate substantially from one leads to a deterioration in the performance of the proposed estimator. These findings justify the use of a data-driven logistic-based $\alpha$ estimator in our simulations, as it adapts well to the observed sample characteristics while maintaining the robustness and accuracy of the Simpson estimator.

## 6.3 Simulation Results

This section evaluates the finite-sample performance of the proposed Simpson-based estimators for the Weitzman overlapping coefficient $\Delta_3$ under four configurations corresponding to large, moderate, small, and very small overlap. Two sample-size settings were considered: (50,50,50) and (100,150,200). The performance measures reported in Table (2) are the average estimate (AV), relative bias (RB), relative root mean squared error (RRMSE), and relative efficiency (EFF) relative to the benchmark estimator $\widehat{\Delta}_3^*$. Since $0 \leq \Delta_3 \leq 1$, values close to one indicate substantial overlap among the three distributions, whereas values near zero reflect limited overlap.

When the true overlap is large ($\Delta_3 = 0.9292$), all estimators perform similarly. For the smaller sample size, the estimators slightly underestimate the true value, with relative bias around −8% and RRMSE close to 0.10. The Simpson-based estimators exhibit efficiencies very close to one, indicating performance nearly identical to the benchmark. As the sample sizes increase, the magnitude of bias decreases and RRMSE declines noticeably, confirming improved precision. Differences among estimators become negligible, showing that in high-overlap settings all approaches are comparably reliable.

For the moderate overlap case ($\Delta_3 = 0.6890$), all estimators display positive bias. In the smaller sample, relative bias is approximately 23%, and RRMSE is around 0.24. Although this upward bias persists even when the sample size increases, the Simpson-based estimators maintain slightly higher efficiency values than the benchmark, particularly the version with $\alpha = 2$. The increase in sample size reduces variability only moderately in this scenario, yet the estimators remain stable and close to one another in overall performance.

When the true overlap is smaller ($\Delta_3 = 0.4688$), bias becomes minimal for all estimators, with relative bias around −3% in small samples and even smaller in larger samples. The RRMSE values decrease substantially as the sample size grows, demonstrating improved precision. In this setting, the Simpson-based estimators consistently outperform the benchmark in terms of efficiency, with gains of about 5–6%. This pattern indicates that the Simpson-rule approach becomes particularly advantageous when the overlap is moderate to small.

The most pronounced differences appear when the true overlap is very small ($\Delta_3 = 0.0735$). Although all estimators exhibit some negative bias in small samples, the Simpson-based estimators achieve markedly lower RRMSE values and substantially higher efficiency. Efficiency exceeds 1.30 and reaches its highest level for the data-driven Simpson estimator. With larger samples, both bias and RRMSE decrease for all estimators, but the efficiency advantage of the Simpson-based methods remains strong. This confirms that the proposed approach is especially effective in low-overlap situations.

A direct comparison of the three Simpson estimators $\widehat{\Delta}_3^{(simp-1)}$ with $\alpha = 1$, $\widehat{\Delta}_3^{(simp-2)}$ with $\alpha = 2$, and $\widehat{\Delta}_3^{(simp-ML)}$, where $\alpha$ is estimated from the data, shows that their numerical results are remarkably close across all scenarios and sample sizes. The differences in AV, RB, and RRMSE are minor and diminish further as the sample size increases. Although the estimator with $\alpha = 2$ occasionally shows a slight efficiency advantage in moderate-overlap cases, the three versions are practically indistinguishable in larger samples. Importantly, the data-driven estimator does not exhibit instability or increased variability, indicating that estimating $\alpha$ adaptively does not compromise performance.

Overall, the simulation results demonstrate that all estimators improve as sample size increases, as evidenced by the consistent reduction in RRMSE. The Simpson-based estimators are at least as efficient as the benchmark and often more efficient, particularly when the true overlap is small. Moreover, the performance of the Simpson-rule approach is largely insensitive to the choice of $\alpha$, whether fixed at 1 or 2 or estimated from the data. These findings confirm the robustness and practical usefulness of the proposed estimators across a wide range of overlap scenarios.

## 7. Conclusion and Discussion

This paper proposed a class of estimators for the Weitzman overlapping coefficient $\Delta_k$, with particular focus on $\Delta_3$, for three independent normal distributions. The proposed methodology is based on the Simpson numerical integration rule and does not impose restrictive assumptions on the parameter space. Three versions of the estimator were developed, differing only in the choice of the tuning parameter $\alpha$: two fixed choices ($\alpha = 1$ and $\alpha = 2$) and a data-driven choice where $\alpha$ is estimated from the sample.

The simulation study demonstrates that all proposed estimators possess desirable finite-sample properties. In particular, the relative root mean squared error decreases consistently as the sample size increases, confirming the consistency of the estimators. When the true overlap $\Delta_3$ is large, all estimators, including the benchmark estimator, perform almost identically, with very small differences in bias and efficiency. This indicates that in high-overlap settings the choice of estimation method is less critical.

More informative differences emerge when the true overlap is moderate or small. In these cases, the Simpson-based estimators generally achieve higher relative efficiency than the benchmark estimator while maintaining comparable bias. The efficiency gains become especially pronounced when $\Delta_3$ is very small, demonstrating that the Simpson-rule approach is particularly well suited for low-overlap situations. This is an important practical advantage, as accurate estimation becomes more challenging when the common support among distributions is limited.

A key finding of this study is that the performance of the Simpson-based estimator is largely insensitive to the choice of $\alpha$. The three proposed versions produce very similar values of AV, RB, RRMSE, and EFF across all scenarios. Although minor differences appear in small samples, these differences diminish as the sample size increases. Importantly, the data-driven estimator, which determines $\alpha$ from the data, does not exhibit instability or inflated variability. This suggests that adaptive selection of $\alpha$ can be employed without sacrificing reliability.

From a practical standpoint, any of the three Simpson estimators may be confidently used. They offer computational simplicity with nearly identical performance. Overall, the proposed Simpson-rule estimators provide a flexible, robust, and efficient approach for estimating the Weitzman overlapping coefficient among three normal distributions. Future research may extend this framework to larger values of $k$, alternative distributional families, or real-data applications to further assess its empirical performance.

**Table (2).** The AV, RB, RRMSE and EFF of the estimators $\hat{\Delta}_3^*$, $\hat{\Delta}_k^{simp}(1)$, $\hat{\Delta}_k^{simp}(2)$ and $\hat{\Delta}_k^{simp}(\hat{\alpha})$ for all Scenarios.

| Scenario (Exact $\Delta_3$) | $(n_1, n_2, n_3)$ | Metrics | $\hat{\Delta}_3^*$ | $\hat{\Delta}_k^{simp}(1)$ | $\hat{\Delta}_k^{simp}(2)$ | $\hat{\Delta}_k^{simp}(\hat{\alpha})$ |
|---|---|---|---|---|---|---|
| $S_1$ $\Delta_3 = 0.929$ | (50,50,50) | AV | 0.85073 | 0.85029 | 0.84800 | 0.85031 |
| | | RB | -0.08425 | -0.08473 | -0.08719 | -0.08471 |
| | | RRMSE | 0.10149 | 0.10156 | 0.10302 | 0.10155 |
| | | EFF | **1.00000** | **0.99888** | **0.97052** | **0.99888** |
| | (100,150,200) | AV | 0.87408 | 0.87370 | 0.87093 | 0.87373 |
| | | RB | -0.05912 | -0.05953 | -0.06251 | -0.05949 |
| | | RRMSE | 0.07673 | 0.07700 | 0.07838 | 0.07699 |
| | | EFF | **1.00000** | **0.99219** | **0.95849** | **0.99219** |
| $S_2$ $\Delta_3 = 0.689$ | (50,50,50) | AV | 0.84738 | 0.84669 | 0.84441 | 0.84672 |
| | | RB | 0.22986 | 0.22887 | 0.22555 | 0.22891 |
| | | RRMSE | 0.24319 | 0.24214 | 0.23861 | 0.24218 |
| | | EFF | **1.00000** | **1.00898** | **1.03885** | **1.00862** |
| | (100,150,200) | AV | 0.87485 | 0.87450 | 0.87165 | 0.87454 |
| | | RB | 0.26974 | 0.26924 | 0.26510 | 0.26929 |
| | | RRMSE | 0.27789 | 0.27735 | 0.27306 | 0.27740 |
| | | EFF | **1.00000** | **1.00383** | **1.03559** | **1.00356** |
| $S_3$ $\Delta_3 = 0.469$ | (50,50,50) | AV | 0.45462 | 0.45375 | 0.45372 | 0.45375 |
| | | RB | -0.0307 | -0.0325 | -0.0326 | -0.0325 |
| | | RRMSE | 0.12102 | 0.11762 | 0.11758 | 0.11759 |
| | | EFF | **1.00000** | **1.05921** | **1.05921** | **1.05921** |
| | (100,150,200) | AV | 0.45879 | 0.45902 | 0.45899 | 0.45903 |
| | | RB | -0.02176 | -0.02128 | -0.02135 | -0.02126 |
| | | RRMSE | 0.08720 | 0.08474 | 0.08471 | 0.08473 |
| | | EFF | **1.00000** | **1.05696** | **1.05696** | **1.05696** |
| $S_4$ $\Delta_3 = 0.074$ | (50,50,50) | AV | 0.06711 | 0.06835 | 0.06863 | 0.06842 |
| | | RB | -0.0931 | -0.07637 | -0.07251 | -0.07544 |
| | | RRMSE | 0.30476 | 0.26544 | 0.26616 | 0.26499 |
| | | EFF | **1.00000** | **1.30769** | **1.30769** | **1.34211** |
| | (100,150,200) | AV | 0.06940 | 0.06919 | 0.07089 | 0.06936 |
| | | RB | -0.06216 | -0.0650 | -0.0421 | -0.0627 |
| | | RRMSE | 0.25685 | 0.22058 | 0.22370 | 0.22117 |
| | | EFF | **1.00000** | **1.33333** | **1.33333** | **1.33333** |

**Conflict of interest:** The author declares that there is no conflict of interest.

**Use of artificial intelligence tools:** The author used ChatGPT to assist in the preparation of this manuscript. The author takes full responsibility for the content of the manuscript.

Montoya, J. A., Figueroa, G. P., and González-Sánchez, D. (2019). Statistical inference for the Weitzman overlapping coefficient in a family of distributions. *Applied Mathematical Modelling, 71*, 558–568.

Mulekar, M. S., and Mishra, S. N. (1994). Overlap coefficient of two normal densities: Equal means case. *Journal of the Japan Statistical Society, 24*, 169–180.

Mulekar, M. S., and Mishra, S. N. (2000). Confidence interval estimation of overlap: Equal means case. *Computational Statistics & Data Analysis, 34*(2), 121–137.

Nunez-Antonio, G., Mendoza, M., Contreras-Cristan, A., Gutierrez-Pena, E., and Mendoza, E. (2018). Bayesian nonparametric inference for the overlap of daily animal activity patterns. *Environmental and Ecological Statistics, 25*(4), 471–494.

Reiser, B., and Faraggi, D. (1999). Confidence intervals for the overlapping coefficient: The normal equal variance case. *Journal of the Royal Statistical Society: Series D (The Statistician), 48*(3), 413–418.

Ridout, M. S., and Linkie, M. (2009). Estimating overlap of daily activity patterns from camera trap data. *Journal of Agricultural, Biological, and Environmental Statistics, 14*(3), 322–337.

Rudin, W. (1976). *Principles of mathematical analysis* (3rd ed.). McGraw-Hill.

Schmid, F., and Schmidt, A. (2006). Nonparametric estimation of the coefficient of overlapping-Theory and empirical application. *Computational Statistics & Data Analysis, 50*(6), 1583–1596.

Van der Vaart, A. W. (1998). *Asymptotic statistics*. Cambridge University Press.

Wang, D., and Tian, L. (2017). Parametric methods for confidence interval estimation of overlap coefficients. *Computational Statistics & Data Analysis, 106*, 12–26.

Wolff, J., Hefner, G., Normann, C., Kaier, K., Binder, H., Hiemke, C., Toto, S., Domschchke, K., Marschollek, M., and Klimke, A. (2021). Polypharmacy and the risk of drug–drug interactions and potentially inappropriate medications in hospital psychiatry. *Pharmacoepidemiology and Drug Safety, 30*(9), 1258–1268.15